\documentclass[twocolumn,doublespacing,showpacs,preprintnumbers,amsmath,amssymb]{revtex4-1}
\usepackage{amssymb}
\usepackage{txfonts}
\usepackage{graphicx}
\usepackage{subfigure}
\usepackage{dcolumn}
\usepackage{bm}
\usepackage{sidecap}
\usepackage[colorlinks,
            linkcolor=blue,
            anchorcolor=blue,
            citecolor=blue
            ]{hyperref}

\begin{document}

\title{User-defined quantum key distribution}

\author{Zhengyu Li$^{1,2 \dag}$}
\author{Yichen Zhang$^{3 \dag}$}
\author{Hong Guo$^{1}$}
\thanks{Corresponding author: hongguo@pku.edu.cn}

\affiliation{$^1$State Key Laboratory of Advanced Optical Communication Systems and Networks, School of Electronics Engineering and Computer Science, Center for Quantum Information Technology, Center for Computational Science and Engineering, Peking University, Beijing 100871, China}

\affiliation{$^2$Central Research Institute, 2012 Labs, Huawei Technoligies Co., Ltd, Shenzhen, Guangdong, China}

\affiliation{$^3$State Key Laboratory of Information Photonics and Optical Communications, Beijing University of Posts and Telecommunications, Beijing 100876, China }

\affiliation{$^\dag$These authors contribute equally to this work.}

\date{\today}

\begin{abstract}
Quantum key distribution (QKD) provides secure keys resistant to code-breaking quantum computers. As headed towards commercial application, it is crucial to guarantee the practical security of QKD systems. However, the difficulty of security proof limits the flexibility of protocol proposals, which may not fulfill with real application requirements. Here we show a protocol design framework that allows one to securely construct the protocol using arbitrary non-orthogonal states. Multi-mode entangled source is virtually introduced for the security analysis, while coherent measurement is used to provide raw data. This `arbitrary' feature reverses the traditional protocol-decide-the-system working style, such that the protocol design now can follow what the system generates. We show a valuable showcase, which not only solves the security challenge of discrete-modulated coherent states, but also achieves high performance with no more than 256 coherent states. Our findings lower the requirement for system venders with off-the-shell devices, thus will promote the commercialization of QKD.
\end{abstract}

\maketitle

BB84 protocol~\cite{BB84_1984} started the era of quantum cryptography, among which quantum key distribution (QKD)~\cite{Gisin_RevModPhys_2002,Scarani_RevModPhys_2009,Lo_NaturePhoton_2014} is the most applicable technology, providing physical-layer protection of information transmission through secure distribution of private keys. For cost-effective implementation, a practical system usually carries out the prepare-and-measure (PM) scheme of a QKD protocol, in which non-orthogonal states are randomly prepared by Alice (the sender), and transmitted to Bob (the receiver), who will measure the states with either single-photon detection or coherent measurement (homodyne or heterodyne detection)~\cite{Ralph_PhysRevA_1999,Grosshans_PhysRevLett_2002,Weedbrook_PhysRevLett_2004,Patron_PhysRevLett_2009}. Systems with coherent detectors are more attractive to commercial companies, due to its room-temperature operation feature and the compatibility with mature product chain of telecommunication. Protocols with coherent measurement usually encode key information on quadratures of a quantum optical state, which are usually called continuous variable (CV) protocols~\cite{Braunstein_RevModPhys_2005, Weedbrook_RevModPhys_2012, Diamanti_Entropy_2015}.

The most influential CV protocol is GG02 protocol using Gaussian modulated coherent states~\cite{Grosshans_PhysRevLett_2002,Grosshans_Nature_2003}. It later evolves to various Gaussian protocols~\cite{Weedbrook_PhysRevLett_2004,Patron_PhysRevLett_2009,Pirandola_NaturePhys_2008} with theoretical security proof~\cite{Acin_PhysRevLett_2006,Patron_PhysRevLett_2006,Furrer_PhysRevLett_2012,Leverrier_PhysRevLett_2015,Leverrier_PhysRevLett_2017}, outperforming other CV protocols. To maintain the practical security~\cite{Scarani_RevModPhys_2009} of a protocol running in a system, the system should fulfill the theoretical assumptions in security proof.  However, even the most state-of-the-art components cannot remove all the theory-experiment mismatches, for instance, the continuous modulation of Gaussian protocols can never be achieved with finite resolution digital-to-analog-convertor (DAC)~\cite{Paul_PhysRevA_2012}. These mismatches also are one of the motivations for the exploration of CV protocols using discrete modulation~\cite{Zhao_PhysRevA_2009,Weedbrook_PhysRevA_2018,Leverrier_PhysRevLett_2009}, but their performances are not promising as Gaussian protocols. Therefore, it's desired for such a protocol that it is adjustable according to practical systems.

Here we move one step forward, proposing a new CV protocol design framework, which allows one to construct the protocol using arbitrary non-orthogonal states with rigorous security analysis. Numerous protocols can be proposed by choosing different non-orthogonal states, which can be discretely or continuously distributed, and can be pure or mixed. This `arbitrary' feature makes the protocol design can be customized by any system vendor according to what they can actually manufacture.

The framework contains two duel schemes, one is the PM scheme, and the other is the entanglement-based (EB) scheme~\cite{Grosshans_QIC_2003}, which is the core design of our framework. The main idea is that Alice uses multi-mode entangled state as the source, and conducts positive-operator valued measures (POVMs) and coherent measurements on different modes. The results of POVMs correspond to the key information in Alice's side, and decide which state is sent out. The results of coherent measurements are used to estimate the correlation between Alice and Bob, through which the lower bound of the secret key rate can be calculated.

Let us explain our framework using schematics in Fig. 1. Due to finite resolution of devices, discrete modulation is always the case in practical applications, therefore we describe our framework in discrete form. The PM scheme of our framework is quite similar as a general QKD protocol. There are $n \left( { \ge 2} \right)$ different non-orthogonal states $\left\{ {\rho _{{B_0}}^1,\rho _{{B_0}}^2...\rho _{{B_0}}^n} \right\}$ that Alice could possibly send to Bob with non-zero probabilities $\left\{ {{p_1},{p_2}...{p_n}} \right\}$. For each time, which state will be sent is decided by the first random number (a complex number or a vector) generated by a quantum random number generator (QRNG). Bob measures the received state ${\rho _B}$ with coherent measurement, and then they do the post-processing~\cite{Bennett_IEEE_1995,Renner_conference_2005}. The difference is, Alice additionally needs a second sequence for the security analysis (see explanation later).

\begin{figure}[t]
\centerline{\includegraphics[width=0.47\textwidth]{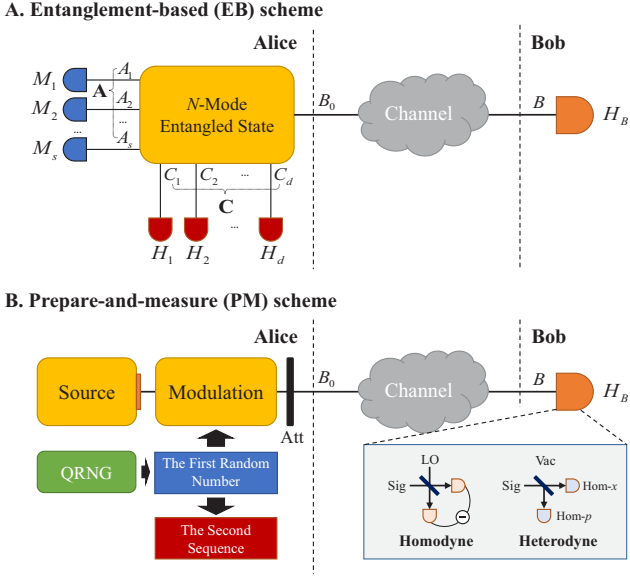}}
\caption{\textbf{Schematics for EB and PM schemes of our framework.} (A) EB scheme. The entangled source is an $N$-mode state, among which there are $s\left( { \ge 1} \right)$ modes ${A_1}{A_2}...{A_s}$ measured by some POVMs, and other $d\left( { \ge 1} \right)$ modes ${C_1}{C_2}...{C_d}$ will be measured by heterodyne detectors. Mode ${B_0}$ is sent to Bob. (B) PM scheme. The state generation is accomplished by modulating the light source, followed with a strong attention. The modulation is controlled by the first random number (or a random vector) which can be acquired through a QRNG. The second sequence is decided by the first random number. Bob's measurement can be either homodyne (Hom) or heterodyne (Het) detection. Att: Attenuator. LO: Local oscillator. Vac: Vacuum. Hom-$x(p)$: homodyne detection for $x(p)$ quadrature. }\label{fig1}
\end{figure}

The equivalence of the EB scheme lies in the design of Alice's entangled source and measurements. The entangled source $\left| {{\psi _{{\bf{AC}}{B_0}}}} \right\rangle $ is an $N$-mode purification of the mixed state ${\rho _{{B_0}}} = \sum\nolimits_{i = 1}^n {{p_i}\rho _{{B_0}}^i} $, in which the subscript ${\bf{A}}$ represents $s$ modes ${A_1}{A_2}...{A_s}$, ${\bf{C}}$ represents $d$ modes ${C_1}{C_2}...{C_d}$, and $s + d + 1 = N,\,s \ge 1,\,d \ge 1$. Alice keeps modes ${\bf{A}}$ and ${\bf{C}}$, while sends mode ${B_0}$ to Bob. The measurements for modes $\bf{A}$ are POVMs, with the results recorded as ${M_{\bf{A}}} = \left( {{m_1},{m_2},...{m_s}} \right)$; and the measurements for modes ${\bf{C}}$ are heterodyne measurements, with the results recorded as ${H_{\bf{C}}} = \left( {{h_1},{h_2}...{h_d}} \right)$. We require that these measurements will project mode ${B_0}$ onto a state $\rho _{{B_0}}^{{M_{\bf{A}}},{H_{\bf{C}}}} \in \left\{ {\rho _{{B_0}}^1,\rho _{{B_0}}^2...\rho _{{B_0}}^n} \right\}$. After sending the state $\rho _{{B_0}}^{{M_{\bf{A}}},{H_{\bf{C}}}}$ to Bob, the rest are the same as the PM scheme.

To show the validity of our framework for arbitrary non-orthogonal states, we first give a sufficient condition to find such an $N$-mode purification. It is that which state is sent to Bob is only decided by the POVMs results ${M_{\bf{A}}}$. This means the sub-state of modes ${\bf{C}}$ and ${B_0}$ conditioned on the results ${M_{\bf{A}}}$ is a product state, $\rho _{{\bf{C}}{B_0}}^{{M_{\bf{A}}}} = \rho _{\bf{C}}^{{M_{\bf{A}}}} \otimes \rho _{{B_0}}^{{M_{\bf{A}}}}$. Then among the purifications of such mixed state ${\rho _{{\bf{C}}{B_0}}} = \sum\nolimits_{i = 1}^n {{p_i}\rho _{\bf{C}}^i \otimes \rho _{{B_0}}^i} $, the entangled source $\left| {{\psi _{{\bf{AC}}{B_0}}}} \right\rangle $ and the corresponding POVMs for modes ${\bf{A}}$ can always be found. This sub-state product feature is also the necessary condition if the non-orthogonal states are coherent states.

\begin{figure}[b]
\centerline{\includegraphics[width=0.47\textwidth]{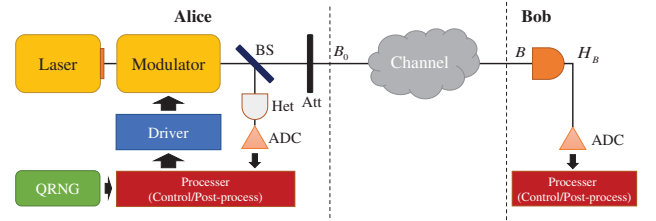}}
\caption{\textbf{State preparation calibration.} In a coherent-state system, to check what state is really generated, a beamsplitter (BS) is inserted between the modulator and the attenuator. One output of the BS will be measured by a heterodyne detector (Het). The measurement result will be sampled and sent to the processor (used for system control and post-processing), in which the relationship between the data and the modulation result is processed. ADC: Analog-to-Digital Convertor.}\label{fig2}
\end{figure}

Second, we explain how to calculate the secret key rate. Here we restrict to the reverse reconciliation and asymptotic case~\cite{Grosshans_Nature_2003,Devetak_Proc_2005}, which is the base for other cases. The key part is to evaluate the Holevo information~\cite{Holevo_Probl_1973}~$S\left( {{H_B}:E} \right)$ between Bob's data and the quantum adversary, whose upper bound $S_{BE}^G$ can be got through the covariance matrix ${\gamma _{{\bf{AC}}B}}$ thanks to the Gaussian state extramelity theorem~\cite{Wolf_PhysRevLett_2006,Patron_PhysRevLett_2006}. However, the POVMs for modes $\bf{A}$ make ${\gamma _{{\bf{AC}}B}}$ incomplete. More specifically, if ${\gamma _{{\bf{AC}}B}}$ is expressed in the form of several sub-matrices,
\begin{equation}
{\gamma _{{\bf{AC}}B}} = \left( {\begin{array}{*{20}{c}}
{{\gamma _{\bf{A}}}}&{{\phi _{{\bf{AC}}}}}&{{\kappa _{{\bf{A}}B}}}\\
{\phi _{{\bf{AC}}}^T}&{{\gamma _{\bf{C}}}}&{{\phi _{{\bf{C}}B}}}\\
{\kappa _{{\bf{A}}B}^T}&{\phi _{{\bf{C}}B}^T}&{{\gamma _B}}
\end{array}} \right)
\end{equation}
then the covariance term ${\kappa _{{\bf{A}}B}}$ is unknown. For other terms, ${\gamma _{\bf{A}}},{\gamma _{\bf{C}}}$, and ${\phi _{{\bf{AC}}}}$ can be theoretically calculated, and ${\phi _{{\bf{C}}B}},{\gamma _B}$ can be estimated through the measured data. Now $S_{BE}^G$ becomes a function of an unknown variable ${\kappa _{{\bf{A}}B}}$. Nevertheless, the uncertainty principle puts a constraint on the covariance matrix of a physical state~\cite{Weedbrook_RevModPhys_2012}, which limits the possible value of ${\kappa _{{\bf{A}}B}}$ to a set ${S_\kappa }$. If we denote $\phi _{{\bf{A}}B}^{Eve}$ as the real eavesdropping induced ${\kappa _{{\bf{A}}B}}$, then $\phi _{{\bf{A}}B}^{Eve} \in {S_\kappa }$. Therefore, by finding the maximum $S_{BE}^G\left( {{\kappa _{{\bf{A}}B}}} \right)$ through traversing the set ${S_\kappa }$, we can define the secret key rate as
\begin{equation}
{K_R} = \beta I\left( {{M_{\bf{A}}}:{H_B}} \right) - \mathop {\sup }\limits_{{\kappa _{{\bf{A}}B}} \in {S_\kappa }} S_{BE}^G\left( {{\kappa _{{\bf{A}}B}}} \right),
\end{equation}
where $I\left( {{M_{\bf{A}}}:{H_B}} \right)$ is the classical mutual information, and $\beta $ is the reconciliation efficiency.

Now we can explain what the second sequence in the PM scheme is. Originally, it should be the measurements results ${H_{\bf{C}}}$, which can be simulated by a QRNG since all modes ${\bf{C}}$ are kept in Alice's side. If further exploiting the product feature of $\rho _{{\bf{C}}{B_0}}^{{M_{\bf{A}}}}$, the estimation of ${\phi _{{\bf{C}}B}}$ requires only the mean values of quadratures for the sub-state $\rho _{\bf{C}}^{{M_{\bf{A}}}}$. Then a simpler form of the second sequence is $\left\{ {\bar x_{{C_1}}^{{M_{\bf{A}}}},\bar p_{{C_1}}^{{M_{\bf{A}}}},...,\bar x_{{C_d}}^{{M_{\bf{A}}}},\bar p_{{C_d}}^{{M_{\bf{A}}}}} \right\}$, where $\bar x_{{C_i}}^{{M_{\bf{A}}}} = {\rm{T}}{{\rm{r}}_{\bf{C}}}\left( {{{\hat x}_{{C_i}}}\rho _{\bf{C}}^{{M_{\bf{A}}}}} \right)$, and $\bar p_{{C_i}}^{{M_{\bf{A}}}} = {\rm{T}}{{\rm{r}}_{\bf{C}}}\left( {{{\hat p}_{{C_i}}}\rho _{\bf{C}}^{{M_{\bf{A}}}}} \right)$. This simple form can be realized digitally in the processer, since it's decided by the first random number, not independently random. Therefore, the PM scheme has no change in hardware comparing to the existing CV-QKD system.

System vendors, as the direct user of protocols, used to build the system following the instruction of a protocol. Now in contrast to this tradition, the protocol can be customized following a practical system. A vendor can start with checking what states their system can generate, then set the probability of sending each state. For the rest, one can follow our framework to find a proper $N$-mode purification, and the secret key rate formula can be got.

Protocols using coherent states are usually the choice of vendors due to the low-cost laser source. The state is generated by modulating the laser with an intensity modulator (IM) and a phase modulator (PM) or a quadrature-phase shift keying (QPSK) modulator, followed by a strong attention. Such modulation using off-the-shell devices usually suffers problems as discretization, non-linearity and noise. To check what the state is actually generated, one can use an additional measurement structure, shown in Fig. 2. The modulated light passes a beamsplitter before entering the attenuator, and a large portion of it goes to a heterodyne detector. Then the modulation result can be read-out with high signal-to-noise ratio (SNR), since the noise figure of classical detectors performs well in the bandwidth of a QKD system (usually less than $1$GHz). This step can be a pre-calibration procedure, or a continuous feedback during the whole running time. Once the map between Alice's data and its real modulation result is set up, it can also be used to compensate the modulation error. Only small deviation remains.

\begin{figure}[t]
\centerline{\includegraphics[width=0.47\textwidth]{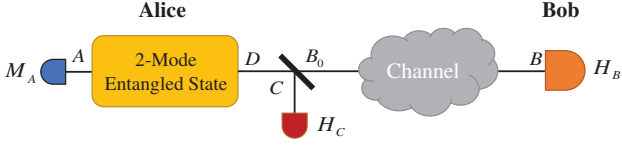}}
\caption{\textbf{The specific three-mode protocol for coherent states.} The three-mode entangled source is acquired through a two-mode entangled state $\left| {{\psi _{AD}}} \right\rangle $. Mode $D$ interacts with the vacuum on a beamsplitter, then the two output modes are $C$ and $B_0$. }\label{fig3}
\end{figure}

We found an effective way to build the EB scheme for the case using coherent states. We choose three-mode entangled source $\left| {{\psi _{AC{B_0}}}} \right\rangle $ for simplicity, and our design principle is to maximize the correlation between modes $C$ and $B_0$, which can limit the eavesdropper. This leads to the choice for each $\rho _C^i$ that it's also a coherent state $\left| {\alpha _C^i} \right\rangle $ with the mean value linearly dependent on $\left| {\alpha _{{B_0}}^i} \right\rangle $. Following this way, one only needs to find a two-mode entangled state $\left| {{\psi _{AD}}} \right\rangle $, then mode $D$ passing a beamsplitter will result in the $\left| {{\psi _{AC{B_0}}}} \right\rangle $, as shown in Fig. 3.

\begin{figure}[t]
\centerline{\includegraphics[width=0.48\textwidth]{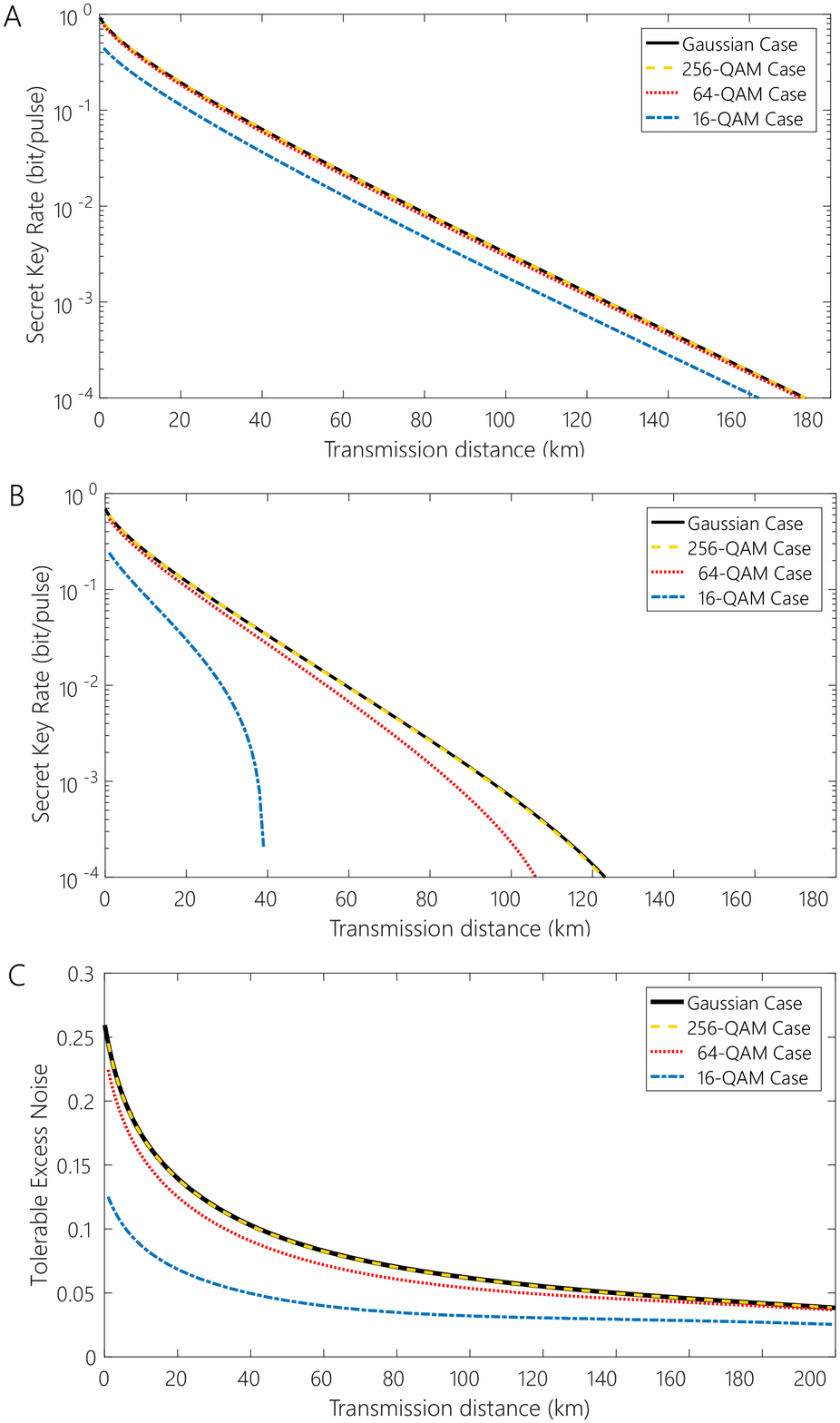}}
\caption{\textbf{Secret key rates and tolerable excess noises.} (A) Secret key rate for low channel noise case ($\epsilon_C=0.01$). (B) Secret key rate for high channel noise case ($\epsilon_C=0.05$). (C) Tolerable excess noise. The black solid line represents the ideal Gaussian case, the blue dash-dotted line represents 16-QAM case, the red dotted line represents 64-QAM case, the yellow dashed line represents 256-QAM case. Simulation details are explained in the supplementary information.}\label{fig4}
\end{figure}

To achieve the same high performance as ideal Gaussian protocols, our framework can reduce the necessary number of used coherent states to no more than 256. Comparing to generating around 1 million coherent states in Gaussian protocols, requested to suppress the theory-experiment mismatch~\cite{Paul_PhysRevA_2012}, this will greatly reduce the complexity of state preparation. Now only 4-bits resolution for each quadrature's modulation is required, which means the modulation noise is negligible, considering the fact that the equivalent-number-of-bit (ENOB) for an off-the-shell DAC can usually reach higher than 10 bits. Different constellation of $\left| {\alpha _{{B_0}}^i} \right\rangle $ will influence the protocol's performance. We show some performance simulations of standard quadrature-amplitude-modulation (QAM) with different number of states in Fig. 4, which is commonly used constellation in classical telecommunication. One can find that with proper settings, 256-QAM can reach the performance almost the same as ideal Gaussian case. And for the low noise case, the number of coherent states can be further reduced to 64, or even lower as 16 for short range. Small-deviation non-standard QAMs, which may happen due to the uncompensated modulation non-linearity, have the similar performances.

Other constellation maps can also be introduced, and run on the same hardware. The switch among pre-set or freshly user-defined constellation maps can be actively controlled by customers, through software-defined manner. This complies the trend of telecommunication network. Combined with the simple modulation and allowance for using off-the-shell devices, our framework will promote the commercialization of QKD.

We thank C. Su, L. Lu, Y. Zou, Y. Cai and B. Xu for discussions. This work was supported by the National Natural Science Foundation under Grant 61531003.\\ \\


\section*{Appendix A: Secret key rate}
Here we explain our derivation of the secret key rate formula.
The state-of-the-art security analysis method is deriving the secret key length formula in the finite-size regime under the universal composable framework (UCF)~\cite{Furrer_PhysRevLett_2012,Leverrier_PhysRevLett_2015,Leverrier_PhysRevLett_2017}. First, one needs to reduce the full formula (quantized by the smooth min-entropy) to a lower bound, which usually is the asymptotic secret key rate with modification terms related to the block size. Then derive a lower bound of the asymptotic secret key rate, which should be calculable through only the measured data. The first reduction relies on several theoretical theorems, differing for different entangled states and measurements used in the protocol, and this is an open question for our framework. Therefore, here we focus on the asymptotic secret key rate formula, and discuss the reverse reconciliation case.

A generally used secret key rate for the asymptotic case is the Devetak-Winter formula~\cite{Devetak_Proc_2005},
\begin{equation}
K = \beta I\left( {{M_{\bf{A}}}:{H_B}} \right) - S\left( {{H_B}:E} \right),
\end{equation}
where $I\left( {{M_{\bf{A}}}:{H_B}} \right)$ is the classical mutual information between Alice and Bob, $\beta $ is the classical reconciliation efficiency, and $S\left( {{H_B}:E} \right)$ is the Holevo information between Bob's data and the adversary~\cite{Holevo_Probl_1973}. Usually, $S\left( {{H_B}:E} \right)$ can be replaced by any of its upper bounds $\bar S\left( {{H_B}:E} \right)$, among which the Gaussian state extramelity theorem~\cite{Wolf_PhysRevLett_2006,Patron_PhysRevLett_2006} induced upper bound $S_{BE}^G$ is the most commonly used case. Because its calculation only relies on the covariance matrix ${\gamma _{{\bf{AC}}B}}$, which can be estimated through the experimental data.

The covariance matrix $\gamma$ of a \emph{N}-mode state $ \hat \rho_N$ is defined as~\cite{Weedbrook_RevModPhys_2012},
\begin{equation}
\gamma_{ij}:=\frac{1}{2} \left \langle \left\{ \Delta \hat r_i,\Delta \hat r_j \right\} \right\rangle,
\end{equation}
where ${ \bf{\hat r}} = \left\{ \hat x_1, \hat p_1,..., \hat x_N, \hat p_N   \right\}$, $\left \langle \hat r_i\right\rangle = {\rm{Tr}}\left( \hat r_i \hat \rho_N \right)$, and $\Delta \hat r_i = \hat r_i -  \left \langle \hat r_i\right\rangle$. Suppose ${\gamma _{{\bf{AC}}B}}$ is the covariance matrix of the state ${\rho _{{\bf{AC}}B}}$, which is the state after mode ${B_0}$ of the entangled source $\left| {{\psi _{{\bf{AC}}{B_0}}}} \right\rangle $ arriving at Bob's side through the channel.
It can be represented using several sub-matrices,
\begin{equation}
{\gamma _{{\bf{AC}}B}} = \left( {\begin{array}{*{20}{c}}
   {{\gamma _{\bf{A}}}} & {{\phi _{{\bf{AC}}}}} & {{\kappa _{{\bf{A}}B}}}  \\
   {\phi _{{\bf{AC}}}^T} & {{\gamma _{\bf{C}}}} & {{\phi _{{\bf{C}}B}}}  \\
   {\kappa _{{\bf{A}}B}^T} & {\phi _{{\bf{C}}B}^T} & {{\gamma _B}}  \\
\end{array}} \right)
\end{equation}
where ${\gamma _{\bf{A}}}$, ${\gamma _{\bf{C}}}$ and ${\gamma _B}$ are covariance matrices for modes $\bf{A}$, $\bf{C}$ and $B$, and ${\phi _{{\bf{AC}}}}$, ${\phi _{{\bf{C}}B}}$ and ${\kappa _{{\bf{A}}B}}$ are covariance terms between different modes.

Among all these sub-matrices, ${\gamma _{\bf{A}}}$, ${\gamma _{\bf{C}}}$ and ${\phi _{{\bf{AC}}}}$ can be directly calculated from $\left| {{\psi _{{\bf{AC}}{B_0}}}} \right\rangle $, since modes ${\bf{A}}$ and ${\bf{C}}$ are kept in Alice's side. ${\phi _{{\bf{C}}B}}$ and ${\gamma _B}$ can be estimated after Alice and Bob randomly sharing part of their coherent measurement results. The only unknown sub-matrix is ${\kappa _{{\bf{A}}B}}$, since the measurements for modes $\bf{A}$ are not coherent measurements now.

Nevertheless, the covariance matrix ${\gamma _N}$ for a $N$-mode state is constrained by the uncertainty principle~\cite{Weedbrook_RevModPhys_2012}, which is
\begin{equation}
{\gamma _N} + i{\Omega _N} \ge 0,
 \end{equation}
where ${\Omega _N} = diag\left( {\omega_1 ,\omega_2 ,...,\omega_N} \right)$, and
\begin{equation}
\omega_1 =...= \omega_N = \omega  = \left( {\begin{array}{*{20}{c}}
   0 & 1  \\
   { - 1} & 0  \\
\end{array}} \right).
\end{equation}
We denote $S_\kappa ^\psi $ as the set of all ${\kappa _{{\bf{A}}B}}$ satisfying this constraint for ${\gamma _{{\bf{AC}}B}}$, which is
\begin{equation}
S_\kappa ^\psi  = \left\{ {\left. {{\phi _{{\bf{A}}B}}} \right|{\gamma _{{\bf{AC}}B}}\left[ {{\kappa _{{\bf{A}}B}} = {\phi _{{\bf{A}}B}}} \right] + i{\Omega _N} \ge 0} \right\}.
\end{equation}
If $\phi _{{\bf{A}}B}^{Eve}$ is the real eavesdropping induced ${\kappa _{{\bf{A}}B}}$, then $\phi _{{\bf{A}}B}^{Eve} \in {S_\kappa^\psi }$.
It can be understood that $S_{BE}^G$ is a function of ${\kappa _{{\bf{A}}B}}$ now. Then by traversing the set $S_\kappa ^\psi $ for all possible ${\kappa _{{\bf{A}}B}}$, we can find the maximum of $S_{BE}^G$. Then the secret key rate can be wrote as
\begin{equation}
{K_R} = \beta I\left( {{M_{\bf{A}}}:{H_B}} \right) - \mathop {\sup }\limits_{{\kappa _{{\bf{A}}B}} \in {S_\kappa ^\psi }} S_{BE}^G\left({\kappa _{{\bf{A}}B}}\right).\label{LowestSKR}
\end{equation}

Next we briefly introduce the calculation method for each term. For $\beta I\left(M_{\bf{A}}:H_B\right)$, it can be expressed as $\beta I\left(M_{\bf{A}}:H_B\right) = H\left(H_B\right) - leak_{EC}$, in which $H\left(H_B\right)$ is the Shannon entropy of Bob's measurement results, and $leak_{EC}$ represents the information that Bob sends to Alice for the data reconciliation. Both these terms can be got from measured data and the error correction step. The reason that we usually separate $\beta I\left(M_{\bf{A}}:H_B\right)$ into $\beta$ and $I\left(M_{\bf{A}}:H_B\right)$ in theoretical study is this helps numerical simulation, which is used to evaluate the performance of a protocol. Given the channel model (usually required to fit the experimental environment), the probability $p\left(H_B|i\right)$ can be got from the model, which is the probability of getting Bob's measurement result $H_B$ given Alice sending the state $\rho_{B_0}^{i}$. And the overall probability will be $p\left(H_B\right) = \sum_{i=1}^{n} p_i p\left(H_B|i\right)$. If $H_B$ is a discrete variable quantized from the measurement result, then
\begin{equation}
\begin{array}{l}
  I\left(M_{\bf{A}}:H_B\right)  \\
=-\sum\limits_{H_B}p\left(H_B\right){\mathrm{log}}_2{p\left(H_B\right)} + \sum\limits_{i=1}^{n}p_i\sum\limits_{H_B}p\left(H_B|i\right) {\mathrm{log}}_2 {p\left(H_B|i\right)}.
\end{array}
\end{equation}
If consider $H_B$ as the continuous variable for some theoretical research, the sum of $H_B$ will be replaced by integration. The reconciliation efficiency $\beta$ can be set according to certain error correction code, for instance, 0.95 is achievable for multi-dimensional reconciliation method in low signal-to-noise regime~\cite{Jouguet_NaturePhoton_2013}.

For $\mathrm{sup}_{\kappa_{{\bf{A}}B}\in S^{\psi}_{\kappa}}~S^G_{BE}\left(\kappa_{{\bf A}B}\right)$, we traverse each $\kappa_{{\bf A}B} \in S^{\psi}_{\kappa}$ to calculate its corresponding $S^G_{BE}\left(\kappa_{{\bf A} B}\right)$ and find the maximal value of them. $S^G_{BE}\left(\kappa_{{\bf A} B}\right)$ can be expressed as $S\left( \rho^G_{{\bf AC}B} | \kappa_{{\bf A}B} \right) - S\left( \rho^G_{{\bf AC}|B} | \kappa_{{\bf A}B} \right)$, in which $S\left( \rho^G_{{\bf AC}B} | \kappa_{{\bf A}B} \right)$ means the von Neumann entropy of a Gaussian state $\rho^G_{{\bf AC}B}$ which has the same covariance matrix as $\gamma_{{\bf AC}B}\left(\kappa_{{\bf A}B}\right)$, and $S\left( \rho^G_{{\bf AC}|B} | \kappa_{{\bf A}B} \right)$ means the von Neumann entropy of a conditional Gaussian state $\rho^G_{{\bf AC}|H_B}$ which has the same covariance matrix as $\gamma_{{\bf AC}|H_B}$, related to Bob's measurement method. The methods to get the $\gamma_{{\bf AC}|H_B}$ from $\gamma_{{\bf AC}B}$, to get the symplectic eigenvalues of each covariance matrix, and to calculate the von Neumann entropy are commonly used in CV-QKD, and can be found in reference~\cite{Patron_PhysRevLett_2006,Weedbrook_RevModPhys_2012}.

\section*{Appendix B: The simple form of the second sequence}
In the covariance matrix ${\gamma _{{\bf{AC}}B}}$, ${\gamma _{\bf{A}}}$, ${\gamma _{\bf{C}}}$ and ${\phi _{{\bf{AC}}}}$ can be theoretically calculated, and ${\gamma _B}$ is estimated only using Bob's data. Only the estimation of ${\phi _{{\bf{C}}B}}$ will use the measurement results of modes \textbf{C}. Naturally, the second sequence should be the measurement results $H_{\bf C}$ for modes $\bf C$, which can be simulated through a quantum random number generator (QRNG). However, the product feature of sub-state $\rho^i_{{\bf C}B_0}$ can help to simplify this.

Let's take the $x$-quadrature of mode ${C_1}$ as an example. After tracing out the other modes of \textbf{C}, the state of modes ${C_1}$ and $B_0$ is ${\rho _{{C_1}{B_0}}} = \sum\nolimits_{i = 1}^n {{p_i}\rho _{{C_1}}^i \otimes \rho _{{B_0}}^i} $, where $\rho _{{C_1}}^i = {{\rm Tr}_{{C_2}...{C_s}}}\left( {\rho _{\bf{C}}^i} \right)$. Then
\begin{eqnarray}
 \left\langle {\Delta {\hat x_{{C_1}}}\Delta {\hat x_{{B_0}}}} \right\rangle
 &=& \sum\limits_{i = 1}^n {{p_i}{{\rm Tr}_{{C_1}{B_0}}}\left[ {\left( {{{\hat x}_{{C_1}}} - {{\bar x}_{{C_1}}}} \right)\left( {{{\hat x}_{{B_0}}} - {{\bar x}_{{B_0}}}} \right)\rho _{{C_1}}^i \otimes \rho _{{B_0}}^i} \right]} \nonumber  \\
  &=& \sum\limits_{i = 1}^n {{p_i}\left( {\bar x_{{C_1}}^i - {{\bar x}_{{C_1}}}} \right)\left( {\bar x_{{B_0}}^i - {{\bar x}_{{B_0}}}} \right)},
\end{eqnarray}
in which each $\bar x_{{C_1}}^i = {{\rm Tr}_{{C_1}}}\left( {{{\hat x}_{{C_1}}}\rho _{{C_1}}^i} \right)$ and ${\bar x_{{C_1}}} = \sum\nolimits_{i = 1}^n {{p_i}\bar x_{{C_1}}^i} $ can be theoretically calculated. Therefore, if the first random number decides that $\rho _{{B_0}}^i$ will be sent, then it's enough to let the second sequence be $\left\{ {\bar x_{{C_1}}^i,\bar p_{{C_1}}^i,...,\bar x_{{C_d}}^i,\bar p_{{C_d}}^i} \right\}$ for the calculation of $\phi_{{\bf C}B}$. 
 This is much simpler than simulating the heterodyne results $H_{\bf C}$ using quantum random numbers.

\section*{Appendix C: The product feature of sub-state $\rho_{{\bf{C}}{B_0}}^{M_{\bf{A}}}$ is necessary}
The EB scheme plays the key role in our framework, in which we need to find a proper entangled source and the measurements in Alice's side. Here we explain one detail of our design solution. Consider the case that one wants to use finite discrete-distributed coherent states as the non-orthogonal source, which is the most significant case for practical implementation. In this case, it is necessary to require that the sub-state $\rho_{{\bf{C}}{B_0}}^{M_{\bf{A}}}$ of modes ${\bf{C}}$ and $B_0$ conditioned on the POVMs results $M_{\bf{A}}$ is a product state.

To explain this necessity, we first prove a lemma which is
\\

\noindent {\textbf{Lemma 1.}} For any two-mode entangled state ${\rho _{AB}}$, if after the heterodyne detection over mode $A$, mode $B$ is projected onto a coherent state, then the number of possibly projected coherent states for mode $B$ is either one or infinite.
\\

We prove this by contradiction. Suppose ${\rho _{AB}}$ satisfies that mode $B$ is a coherent state after the heterodyne detection over mode $A$, and the number $n$ of the possibly projected coherent state is $\infty  > n \ge 2$.

For generality, we assume ${\rho _{AB}}$  is a mixed state. There exists a purification of ${\rho _{AB}}$, which can be expressed as ${\rho _{FAB}} = \sum\nolimits_{i = 1}^n {\sqrt {{p_i}} \left| {\phi _{FA}^i} \right\rangle \left| {\alpha _B^i} \right\rangle } $. The heterodyne detection can be seen as the projection onto a coherent state $\left| {{\alpha _A}} \right\rangle $. We divide the overall phase space for mode $A$ into $n+1$ different sets $\left\{ {{\varsigma _1},{\varsigma _2}...{\varsigma _{n + 1}}} \right\}$, among which any two of them has no overlap. This equals to divide the two-dimensional plane into $n+1$ points sets without overlap. The first $n$ sets correspond to the $n$ different output coherent states of mode $B$. For example, for the heterodyne measurement result $\alpha _A$, if ${{\alpha _A}} \in {\varsigma _i}$, then mode $B$ will be projected onto $\left| {\alpha _B^i} \right\rangle $. And we assume for each element of the first $n$ sets, the probability of getting a corresponding heterodyne result for mode $A$ is non-zero. The last set ${\varsigma _{n + 1}}$ corresponds to the points that will never be the heterodyne result, which means ${\rm Tr}_B\left[ {\left\langle {{\alpha _A}} \right|{\rho _{AB}}\left| {{\alpha _A}} \right\rangle } \right] = 0$, if $ {{\alpha _A}}  \in {\varsigma _{n + 1}}$.

Suppose $1\le k \le n$, and $ {{\alpha _A}}  \in {\varsigma _k}$, then we know
\begin{equation}
\left| {\alpha _B^k} \right\rangle \left\langle {\alpha _B^k} \right| = \sum\nolimits_{ij = 11}^{nn} {{C_{ij}}} \left| {\alpha _B^i} \right\rangle \left\langle {\alpha _B^j} \right|,
\end{equation}
where ${C_{ij}} = \left( {{{\sqrt {{p_i}{p_j}} } \mathord{\left/
 {\vphantom {{\sqrt {{p_i}{p_j}} } {p\left( {{\alpha _A}} \right)}}} \right.
 \kern-\nulldelimiterspace} {p\left( {{\alpha _A}} \right)}}} \right) \cdot {\rm Tr}_F\left[ {\left\langle {{{\alpha _A}}}
 \mathrel{\left | {\vphantom {{{\alpha _A}} {\phi _{FA}^i}}}
 \right. \kern-\nulldelimiterspace}
 {{\phi _{FA}^i}} \right\rangle \left\langle {{\phi _{FA}^j}}
 \mathrel{\left | {\vphantom {{\phi _{FA}^j} {{\alpha _A}}}}
 \right. \kern-\nulldelimiterspace}
 {{{\alpha _A}}} \right\rangle } \right]$, and $p\left( {{\alpha _A}} \right)$ is the probability of getting the measurement result ${\alpha _A}$.

First, we can prove that if $i \ne k$ or $j \ne k$, then ${C_{ij}} = 0$, which means the only non-zero term is ${C_{kk}} = 1$. The intuitive understanding of this is that, there are infinite equations constraining finite variables ${C_{ij}}$. The detailed proof can follow these steps:

1) derive $D\left(-\alpha_B^k\right)\left| {\alpha _B^k} \right\rangle \left\langle {\alpha _B^k} \right|D^{\dag}\left(-\alpha_B^k\right)$, and move the vacuum state term to one side,
\begin{equation}
\left( {1 - {C_{kk}}} \right)\left| 0 \right\rangle \left\langle 0 \right| = \sum\nolimits_{ij = 11,ij \ne kk}^{nn} {{C_{ij}}} \left| {\alpha _B^i - \alpha _B^k} \right\rangle \left\langle {\alpha _B^j - \alpha _B^k} \right|.
\end{equation}

2) calculate the inner product between the Fock state $\left| t \right\rangle $ and the vacuum state,
\begin{equation}
0 = \sum\nolimits_{ij = 11,ij \ne kk}^{nn} {{C_{ij}}} {e^{ - {{{{\left| {{\beta _i}} \right|}^2}} \mathord{\left/
 {\vphantom {{{{\left| {{\beta _i}} \right|}^2}} 2}} \right.
 \kern-\nulldelimiterspace} 2}}}{e^{ - {{{{\left| {{\beta _j}} \right|}^2}} \mathord{\left/
 {\vphantom {{{{\left| {{\beta _j}} \right|}^2}} 2}} \right.
 \kern-\nulldelimiterspace} 2}}}{{\beta _i^t\beta _j^{*t}} \mathord{\left/
 {\vphantom {{\beta _i^t\beta _j^{*t}} {t!}}} \right.
 \kern-\nulldelimiterspace} {t!}},\quad \forall t \ge 1,
\end{equation}
where ${\beta _i} = \alpha _B^i - \alpha _B^k,{\beta _j} = \alpha _B^j - \alpha _B^k$.

If let ${\lambda _{ij}} = {\beta _i}\beta _j^*$, ${d_{ij}} = {C_{ij}}\exp \left[ { - {{\left( {{{\left| {{\beta _i}} \right|}^2} + {{\left| {{\beta _j}} \right|}^2}} \right)} \mathord{\left/
 {\vphantom {{\left( {{{\left| {{\beta _i}} \right|}^2} + {{\left| {{\beta _j}} \right|}^2}} \right)} 2}} \right.
 \kern-\nulldelimiterspace} 2}} \right]$, then the first ${n^2} - 1$ equations can be written in the matrix formula $\Lambda {\bf{D}} = 0$, in which $\Lambda $ is the transport of a Vandermonde matrix with ${n^2} - 1$ different non-zero ${\lambda _{ij}}$, and ${\bf{D}}$ is a vector with ${n^2} - 1$ different ${d_{ij}}$. Since the Vandermonde matrix has the feature of full rank, then the above equations have the only solution that each ${d_{ij}} = 0$, which means $\forall ij \ne kk,{C_{ij}} = 0$, and ${C_{kk}} = 1$.

Second, for any ${\varsigma _k}, 1\le k\le n+1$, we can define a corresponding set $\tau_k$ of coherent states, which is $\tau_k = \left\{ \left| \alpha_A \right\rangle | {~\rm if~} \alpha _A \in \varsigma_k  \right\}$. Then the above discussion will lead to the conclusion that, for each $\tau _k$, there exists at least one state orthogonal to it. First look at the case $1 \le k \le n$, in which ${C_{ii}} = 0$ for any $i = j \ne k$. This means $\forall \left|{\alpha _A}\right\rangle \in {\tau _k}, \left\langle {{\alpha _A}} \right|\rho _A^{ii}\left| {{\alpha _A}} \right\rangle  = 0$, where $\rho _A^{ii} = {\rm Tr}_F\left[ \left| \phi _{FA}^i \right\rangle \left\langle \phi _{FA}^i \right|   \right]$.
Then $\rho_A^{ii}$ is orthogonal to the set $\tau_k$. Second, for the case $k=n+1$, from its definition we know ${\rm Tr}_B\left[ {\left\langle {{\alpha _A}} \right|{\rho _{AB}}\left| {{\alpha _A}} \right\rangle } \right] = 0$, if $\left| {{\alpha _A}} \right\rangle  \in {\varsigma _{n + 1}}$. Then $\rho_A = {\rm {Tr}}_B\left[ \rho_{AB} \right]$ is orthogonal to $\tau_{n+1}$.

However, it can be proved that when $n$ is finite, among all these $n+1$ set $\left\{ {{\tau _1},{\tau _2}...{\tau _{n + 1}}} \right\}$, at least there is one of them being a complete or over-complete set of the Fock state space~\cite{Bargmann_RepMathPhys_1971}, which means no state can be orthogonal to this set. This is contradictory to the previous conclusion. Therefore, for the case $\infty  > n \ge 2$, no such a two-mode entangled state can be found. $\Box$

One can easily generalize this to the $N$-mode entangled state case, which is, for any $N$-mode entangled state ${\rho _{{\bf{A}}B}}$, if after the heterodyne detections for each mode of ${\bf{A}}$, mode $B $ is projected onto a coherent state, then the number of possibly projected coherent states for mode $B$ is either one or infinite.

For our EB scheme, after the POVMs for modes ${\bf{A}}$, the conditioned sub-state $\rho _{{\bf{C}}{B_0}}^{{M_{\bf{A}}}}$ will face the same situation as the above argument. And what we consider is the finite coherent states case, then the number of possibly projected coherent states for mode $B_0$ is only one. This means $\rho _{{\bf{C}}{B_0}}^{{M_{\bf{A}}}}$ is a product state that $\rho _{{\bf{C}}{B_0}}^{{M_{\bf{A}}}} = \rho _{\bf{C}}^{{M_{\bf{A}}}} \otimes \left| {\alpha _{{B_0}}^{{M_{\bf{A}}}}} \right\rangle \left\langle {\alpha _{{B_0}}^{{M_{\bf{A}}}}} \right|$.

The above conclusion shows four facts about our framework: 1) POVM measurements other than heterodyne detection should be introduced; 2) entangled state with more than two modes are necessary; 3) after the POVMs, the conditioned sub-state should be a product state; 4) which coherent state will be sent to Bob is decided by the results of POVMs.

\section*{Appendix D: Three-mode protocol for finite discrete-distributed coherent states}
For a CV system, generating discrete-distributed coherent states is the most practical case, because of the finite resolution for practical devices. The successful application of our framework to this case improves the practical security of CV systems.

Here we will explain some details of our design principle for the discrete-distributed coherent states case. The three-mode entangled source model we use is not only simple-structure, but also highly effective.

\subsection{Two-mode entangled source}
Suppose the source states are $n$ different coherent states ${S_\rho } = \left\{ {\left| {\alpha _{{B_0}}^1} \right\rangle ,\left| {\alpha _{{B_0}}^2} \right\rangle ...\left| {\alpha _{{B_0}}^n} \right\rangle } \right\}$, and the three-mode entangled source is $\left| {{\psi _{AC{B_0}}}} \right\rangle $. From the discussion of section II.~B we know that, the performance of a protocol is mainly decided by the structure of ${\rho _{{\bf{C}}{B_0}}}$. Therefore, our design principle is to let the correlation between modes $C$ and $B_0$ as `high' as possible, which from the covariance matrix pointview we want $\left| {\left\langle {\Delta {\hat x_C}\Delta {\hat x_{{B_0}}}} \right\rangle } \right|$ and $\left| {\left\langle {\Delta {\hat p_C}\Delta {\hat p_{{B_0}}}} \right\rangle } \right|$ to be as large as possible. We note that our design principle is only an example inspired by the experience, and it works well for the quadrature-amplitude modulation (QAM) case. Other design solutions for different modulation cases also worth further investigations.

Suppose the covariance matrix ${\gamma _{AC{B_0}}}$ for $\left| {{\psi _{AC{B_0}}}} \right\rangle $ is of the standard form~\cite{Duan_PhysRevLett_2000,Simon_PhysRevLett_2000}, which means all $\left\langle {\Delta {\hat x_i}\Delta {\hat p_j}} \right\rangle  = 0, \forall i,j$. This can be achieved in the QAM case.

We take $x$-quadure as the example. Denote the mean and the variance of each sub-state $\rho_C^i$ and $\left| \alpha_{B_0}^i \right\rangle$ as
\begin{equation}
\begin{array}{l}
 \bar x_C^i = T{r_C}\left( {{{\hat x}_C}\rho _C^i} \right),V_{C,x}^i = T{r_C}\left( {\hat x_C^2\rho _C^i} \right) - {\left( {\bar x_C^i} \right)^2}, \\
 \bar x_{{B_0}}^i = \left\langle {\alpha _{{B_0}}^i} \right|{{\hat x}_{{B_0}}}\left| {\alpha _{{B_0}}^i} \right\rangle , V_{{B_0},x}^i = 1.
\end{array}
\end{equation}
Then the overall mean values of modes $C$ and $B_0$ are ${\bar x_C} = \sum\nolimits_{i = 1}^n {{p_i}\bar x_C^i} ,{\bar x_{{B_0}}} = \sum\nolimits_{i = 1}^n {{p_i}\bar x_{{B_0}}^i} $, and the variances are
\begin{equation}
\begin{array}{l}
 {V_{C,x}} = \sum\limits_{i = 1}^n {{p_i}V_{C,x}^i}  + \sum\limits_{i = 1}^n {{p_i}{{\left( {\bar x_C^i - {{\bar x}_C}} \right)}^2}} , \\
 {V_{{B_0},x}} = 1 + \sum\limits_{i = 1}^n {{p_i}{{\left( {\bar x_{{B_0}}^i - {{\bar x}_{{B_0}}}} \right)}^2}} .
 \end{array}
\end{equation}
Now look at $\left| {\left\langle {\Delta {\hat x_C}\Delta {\hat x_{{B_0}}}} \right\rangle } \right|$, which is
\begin{eqnarray}
 \left| {\left\langle {\Delta {\hat x_C}\Delta {\hat x_{{B_0}}}} \right\rangle } \right| &=& \left| {\left\langle {\left( {{{\hat x}_C} - {{\bar x}_C}} \right)\left( {{{\hat x}_{{B_0}}} - {{\bar x}_{{B_0}}}} \right)} \right\rangle } \right|  \nonumber \\
  &=& \left| {\sum\limits_{i = 1}^n {{p_i}{{\rm Tr}_{C{B_0}}}\left[ {\left( {{{\hat x}_C} - {{\bar x}_C}} \right)\left( {{{\hat x}_{{B_0}}} - {{\bar x}_{{B_0}}}} \right)\rho _C^i \otimes \rho _{{B_0}}^i} \right]} } \right| \nonumber \\
  &=& \left| {\sum\limits_{i = 1}^n {{p_i}\left( {\bar x_C^i - {{\bar x}_C}} \right)\left( {\bar x_{{B_0}}^i - {{\bar x}_{{B_0}}}} \right)} } \right| \nonumber \\
  &\le& \sqrt {\left( {\sum\limits_{i = 1}^n {{p_i}{{\left( {\bar x_C^i - {{\bar x}_C}} \right)}^2}} } \right)\left( {\sum\limits_{i = 1}^n {{p_i}{{\left( {\bar x_{{B_0}}^i - {{\bar x}_{{B_0}}}} \right)}^2}} } \right)} \nonumber \\
  &=& \sqrt {\left( {{V_{C,x}} - \sum\limits_{i = 1}^n {{p_i}V_{C,x}^i} } \right)\left( {{V_{{B_0},x}} - 1} \right)} .
\end{eqnarray}

\noindent The inequality is due to the Cauchy-Schwarz inequality, in which the equality holds if and only if $\forall i \in \left[ {1,n} \right],{{\left( {\bar x_C^i - {{\bar x}_C}} \right)} \mathord{\left/
 {\vphantom {{\left( {\bar x_C^i - {{\bar x}_C}} \right)} {\left( {\bar x_{{B_0}}^i - {{\bar x}_{{B_0}}}} \right)}}} \right.
 \kern-\nulldelimiterspace} {\left( {\bar x_{{B_0}}^i - {{\bar x}_{{B_0}}}} \right)}} \equiv t$, where $t$ is non-zero. The uncertainty principle tells that for each sub-state, $V_{C,x}^i V_{C,p}^i \ge 1$. If we further assume that $x$ and $p$ are symmetric for each sub-state, then  $V_{C,x}^i \ge 1$, and $\sum\nolimits_{i = 1}^n {{p_i}V_{C,x}^i}  \ge 1$. Thus, to achieve the maximum $\left| {\left\langle {\Delta {\hat x_C}\Delta {\hat x_{{B_0}}}} \right\rangle } \right|$, we need $V_{C,x}^i = 1$. One can find that these two conditions can be both satisfied if each $\rho _{\bf{C}}^i$ is also a coherent state $\left| {\alpha _C^i} \right\rangle $, with the mean value linearly dependent on $\left| {\alpha _{{B_0}}^i} \right\rangle $, which is ${{\forall i \in \left[ {1,n} \right],\alpha _C^i} \mathord{\left/
 {\vphantom {{\forall i \in \left[ {1,n} \right],\alpha _C^i} {\alpha _{{B_0}}^i}}} \right.
 \kern-\nulldelimiterspace} {\alpha _{{B_0}}^i}} = t$. This linear relationship can be got from a beamsplitter model, which is a coherent state $\left| {\beta _D^i} \right\rangle $ with $\beta _D^i = \sqrt {1 + {t^2}} \alpha _{{B_0}}^i$ passes through a beamsplitter with transmittance ${\eta _{BS}} = {1 \mathord{\left/
 {\vphantom {1 {\left( {1 + {t^2}} \right)}}} \right.
 \kern-\nulldelimiterspace} {\left( {1 + {t^2}} \right)}}$.
This means Alice only needs to generate a two-mode entangled state $\left| {{\psi _{AD}}} \right\rangle  = \sum\nolimits_{i = 1}^n {\sqrt {{p_i}} } \left| {R_A^i} \right\rangle \left| {\beta _D^i} \right\rangle $, and then let the mode $D$ pass through a beamsplitter, shown as Fig.~2 in the main context.

As for the design of $\left| {{\psi _{AD}}} \right\rangle $, first, we find a set of orthogonal states $\left\{ {\left| {\theta _D^i} \right\rangle  = \sum\nolimits_{j = 0}^\infty  {{c_{ij}}\left| j \right\rangle } } \right\}$, which can diagonalize the mixed state ${\rho _D}$, such that
\begin{equation}
{\rho _D} = \sum\nolimits_{i = 1}^n {{p_i}} \left| {\beta _D^i} \right\rangle \left\langle {\beta _D^i} \right| = \sum\nolimits_{i = 1}^n {{\upsilon _i}} \left| {\theta _D^i} \right\rangle \left\langle {\theta _D^i} \right|
\end{equation}
Then $\left| {{\psi _{AD}}} \right\rangle $ is defined as
\begin{equation}
\left| {{\psi _{AD}}} \right\rangle  = \sum\nolimits_{i = 1}^n {\sqrt {{\upsilon _i}} } \left| {\varphi _A^i} \right\rangle \left| {\theta _D^i} \right\rangle,
\end{equation}
where $\left| {\varphi _A^i} \right\rangle $ is related to $\left| {\theta _D^i} \right\rangle $ in such a way,
\begin{equation}
\left| {\varphi _A^i} \right\rangle  = \sum\nolimits_{j = 0}^\infty  {{{\left( {{c_{ij}}} \right)}^*}\left| j \right\rangle }.
\end{equation}

\begin{figure}[htpb]
\centerline{\includegraphics[width=0.48\textwidth]{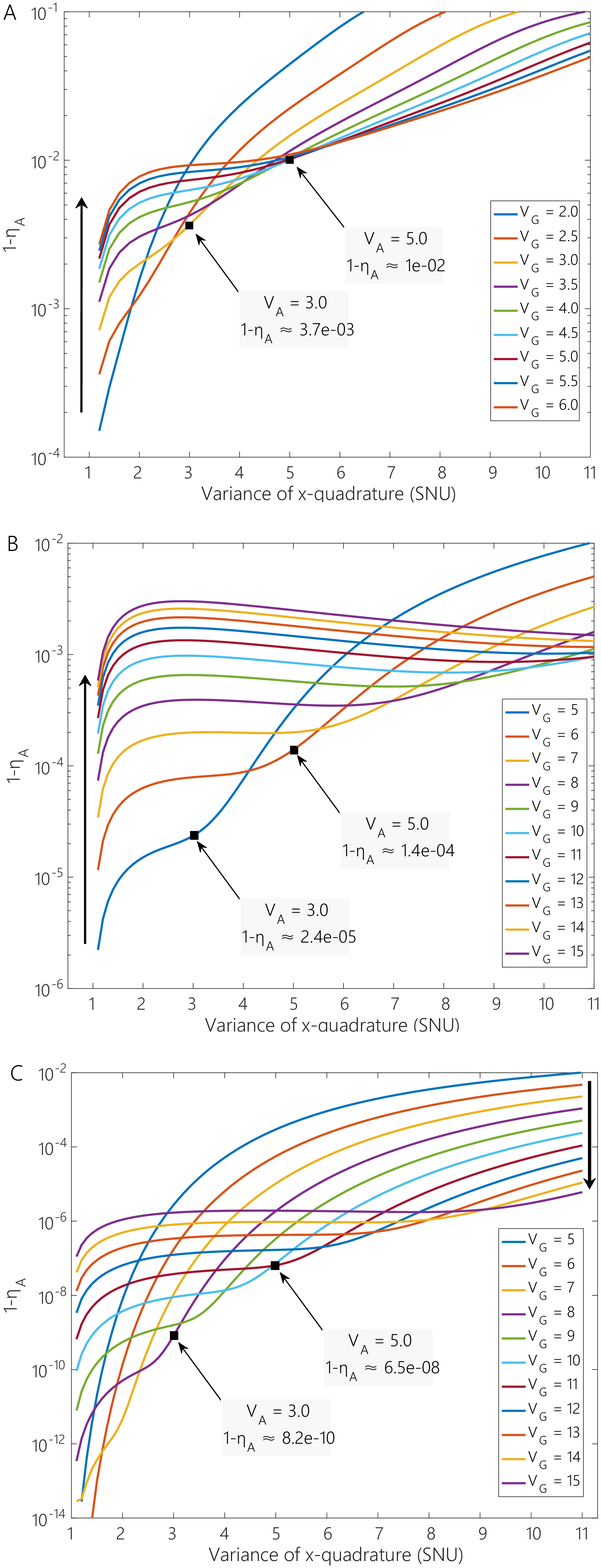}}
\caption{ {\bf $1-\eta_A$ versus $V_A$}.~(A)~16-QAM case, curves following the direction of the arrow (left side bottom-up) correspond to the cases of $V_G = 2.0 \sim 6.0$. (B)~64-QAM case, curves following the direction of the arrow (left side bottom-up) correspond to the cases of $V_G = 5 \sim 15$. (C) 256-QAM case, curves following the direction of the arrow (right side top-down) correspond to the cases of $V_G = 5 \sim 15$. We mark the optimal $V_G$ and the corresponding $1-\eta_A$ for two different conditions $V_A = 3$ and $V_A = 5$.}\label{figS1}
\end{figure}

\subsection{Quadrature-amplitude modulation (QAM)}
We are especially interested in the QAM case, because it's a standard modulation format in the classical coherent communication. Systems running such a modulation format is naturally compatible with current industry chain of electro-optical devices.

In  $n$-QAM ($ n = {L^2}, L$) is positive integer), coherent states are positioned at the cross points of equally-spaced $L$ columns and $L$ rows in the phase space (or classically called constellation map). Suppose the space between each column (or row) is $2r$, then the positions of $n$ coherent states are
\begin{equation}
\left\{ {\forall \mu ,\upsilon  \in \left[ {1,L} \right],{\alpha _{\mu \upsilon }} = \left( {2\mu  - 1 - L} \right)r + i \cdot \left( {2\upsilon  - 1 - L} \right)r} \right\}\nonumber
\end{equation}
It can be verified that for this standard QAM format, the covariance matrix for $\left| {{\psi _{AD}}} \right\rangle $ is of the standard form,
\begin{equation}
{\gamma _{AD}} = \left( {\begin{array}{*{20}{c}}
   {{V_A}I} & {{\phi _{AD}}{\sigma _Z}}  \\
   {{\phi _{AD}}{\sigma _Z}} & {{V_A}I}  \\
\end{array}} \right)\label{CM_AD_std}
\end{equation}
where $I = \left( {\begin{array}{*{20}{c}}
   1 & 0  \\
   0 & 1  \\
\end{array}} \right)$, and ${\sigma _Z} = \left( {\begin{array}{*{20}{c}}
   1 & 0  \\
   0 & { - 1}  \\
\end{array}} \right)$. We know that ${\phi _{AD}} \le \sqrt {V_A^2 - 1} $ due to the uncertainty principle, and the closer ${\phi _{AD}}$ approaches to $\sqrt {V_A^2 - 1} $, the better the protocol performance will be. Thus, for the  $n$-QAM, we need to choose the proper sending probabilities  $\left\{ {{p_1},...,{p_{{n}}}} \right\}$ and the space parameter $r$ to make the $\phi_{AD}$ as large as possible. Since the different $\left\{ {{p_1},...,{p_{{n}}}} \right\}$ and $r$ will result in different ${V_A}$, we introduce a dimensionless parameter ${\eta _A} = {{\phi _{AD}^2} \mathord{\left/
 {\vphantom {{\phi _{AD}^2} {\left( {V_A^2 - 1} \right)}}} \right.
 \kern-\nulldelimiterspace} {\left( {V_A^2 - 1} \right)}}$ to evaluate the closeness of ${\phi _{AD}}$ to $\sqrt {V_A^2 - 1} $ for the small ${V_A}$ region.

Fully optimization of the probabilities $\left\{ {{p_1},...,{p_{{n}}}} \right\}$ is complicated. Here we let them follow a discrete Gaussian distribution: let ${r_0} = 1$ (the unit is the square root of the shot noise unit (SNU)), then the probability $p\left( {{\alpha _{\mu \upsilon }}} \right)$ of sending the state $\left| {{\alpha _{\mu \upsilon }}} \right\rangle $ is
\begin{equation}
p\left( {{\alpha _{\mu \upsilon }}} \right) \propto \exp \left[ { - {{{{\left| {{\alpha _{\mu \upsilon }}\left( {r = {r_0}} \right)} \right|}^2}} \mathord{\left/
 {\vphantom {{{{\left| {{\alpha _{\mu \upsilon }}\left( {r = {r_0}} \right)} \right|}^2}} {\left( {2{V_G}} \right)}}} \right.
 \kern-\nulldelimiterspace} {\left( {2{V_G}} \right)}}} \right].
\end{equation}
This simplifies the probability distribution to only one parameter ${V_G}$.

We numerically calculate the ${\eta _A}$ for $16$-QAM, $64$-QAM and $256$-QAM, with different ${V_G}$ and $r$, to find a relatively optimal combination of $V_A$ and $\eta_A$. Generally speaking, for the small $V_A$ region, the larger the $V_A$ is, the worse the $\eta_A$ is. Fig.~\ref{figS1} shows our simulation result. For $16$-QAM (Fig.~\ref{figS1}(A)), when $V_A = 3$, the optimal choice for $V_G$ is $V_G = 3$, which corresponds to $\eta_A \approx 1-3.7\times10^{-3}$; when $V_A = 5$, the optimal choice for $V_G$ is $V_G = 4.5$, which corresponds to $\eta_A \approx 1-1\times10^{-2}$. For $64$-QAM (Fig.~\ref{figS1}(B)), when $V_A = 3$, the optimal choice for $V_G$ is $V_G = 5$, which corresponds to $\eta_A \approx 1-2.4\times10^{-5}$; and when $V_A = 5$, the optimal choice for $V_G$ is $V_G = 6$, which corresponds to $\eta_A \approx 1 - 1.4\times10^{-4}$. For $256$-QAM (Fig.~\ref{figS1}(C)), when $V_A = 3$, the optimal choice for $V_G$ is $V_G = 8$, which corresponds to $\eta_A \approx 1-8.2\times10^{-10}$; and when $V_A = 5$, the optimal choice for $V_G$ is $V_G = 11$, which corresponds to $\eta_A \approx 1 - 6.5\times10^{-8}$.

From the $\eta_A$-pointview, 256-QAM is almost the ideal Gaussian case, and such a small deviation won't cause a large performance reduction. This is verified by the secret key rate simulation shown in the Fig. 3 of the main context.

In experiment, the two quadratures can be modulated separately, for instance using the QPSK modulator. Thus, for the ${L^2}$-QAM, the resolution of the DAC device for the modulation of one quadrature is $res = {\log _2}L$. Then for $16$-QAM, $res = 2$, $64$-QAM, $res = 3$, and for $256 $-QAM, $res = 4$. DAC devices with such resolutions are off-the-shell and cost-effective.

\subsection{Techniques for the numerical calculation }
Numerical calculation is needed in two parts, in which the first is the calculation of ${\gamma _A}$, ${\gamma _C}$ and ${\phi _{AC}}$, and the second is the searching process.

For the calculation of ${\gamma _A}$, ${\gamma _C}$ and ${\phi _{AC}}$, a simple way is to express every state, e.g. $\left|\alpha^i_{B_0}\right\rangle, \left| {\theta _D^i} \right\rangle$ and $\left| {\phi _D^i} \right\rangle$, in the Fock state basis, and then finish the calculation. One thing needs to be careful with is, to achieve high precision, the number of Fock state bases should be greatly larger than $\sqrt {{V_A}} $. For example, for $V_A = 10$, we choose first 200 Fock states to express a state. This numerical method fits well with the theoretical results for the $4$-QAM case in~\cite{Leverrier_PhysRevLett_2009}.

For the searching process, although the three-mode entangled state model already has the least unknown parameters, the symmetry of the QAM case can further simplify it.

First, we define the standard form of the covariance matrix for the three-mode EB scheme for ${L^2}$-QAM. After Bob sharing part of his measurement results, the sub covariance matrix $\gamma _{CB}$ of modes $C$ and $B$ can be transformed to the standard form $\gamma _{CB}^{std}$~\cite{Duan_PhysRevLett_2000,Simon_PhysRevLett_2000}, which is
\begin{equation}
\gamma _{CB}^{std} = \left( {\begin{array}{*{20}{c}}
   {\left[ {\left( {1 - {\eta _{BS}}} \right)\left( {{V_A} - 1} \right) + 1} \right]I} & {\left( {\begin{array}{*{20}{c}}
   {{\phi _x}} & 0  \\
   0 & {{\phi _p}}  \\
\end{array}} \right)}  \\
   {\left( {\begin{array}{*{20}{c}}
   {{\phi _x}} & 0  \\
   0 & {{\phi _p}}  \\
\end{array}} \right)} & {{V_B}I}  \\
\end{array}} \right),
\end{equation}
through two local unitary operators over modes $C$ and $B$ with corresponding symplectic matrices ${S_C}$ and ${S_B}$. ${\phi _x}$ and ${\phi _p}$ may not be equal. Then one can find a unitary operator over mode $A$ with the corresponding symplectic matrix ${S_A}$, such that ${S_A}{\sigma _Z}S_C^T = {\sigma _Z}$. Therefore, the covariance matrix ${\gamma _{ACB}}$ can be `standardized' by these three operators $S = {S_A} \oplus {S_C} \oplus {S_B}$ :
\begin{widetext}
\begin{equation}
\begin{array}{l}
 {\gamma _{ACB}^{std} = S{\gamma _{ACB}}{S^T}}
 = {  \left( {\begin{array}{*{20}{c}}
   {{V_A}I} & { - \sqrt {\left( {1 - {\eta _{BS}}} \right){\eta _A}\left( {V_A^2 - 1} \right)} {\sigma _Z}} & {{\kappa '_{AB}}}  \\
   { - \sqrt {\left( {1 - {\eta _{BS}}} \right){\eta _A}\left( {V_A^2 - 1} \right)} {\sigma _Z}} & {\left[ {\left( {1 - {\eta _{BS}}} \right)\left( {{V_A} - 1} \right) + 1} \right]I} & {\left( {\begin{array}{*{20}{c}}
   {{\phi _x}} & 0  \\
   0 & {{\phi _p}}  \\
\end{array}} \right)}  \\
   {{\kappa '_{AB}}} & {\left( {\begin{array}{*{20}{c}}
   {{\phi _x}} & 0  \\
   0 & {{\phi _p}}  \\
\end{array}} \right)} & {{V_B}I}  \\
\end{array}} \right)}
 \end{array}
\end{equation}
where ${\kappa '_{AB}} = {S_A}{\kappa _{AB}}S_B^T = \left( {\begin{array}{*{20}{c}}
   {{\kappa _{11}}} & {{\kappa _{12}}}  \\
   {{\kappa _{21}}} & {{\kappa _{22}}}  \\
\end{array}} \right)$ is still unknown.
\end{widetext}

For this standard form, it is found that, no matter Bob uses heterodyne or homodyne detection, the secret key rate for the case $\gamma _{ACB}^{std}\left( {{\kappa _{11}},{\kappa _{12}},{\kappa _{21}},{\kappa _{22}}} \right)$ (denote this state as $\rho _{ACB}^{\left(  +  \right)}$) and the case $\gamma _{ACB}^{std}\left( {{\kappa _{11}}, - {\kappa _{12}}, - {\kappa _{21}},{\kappa _{22}}} \right)$ (denote this state as $\rho _{ACB}^{\left(  -  \right)}$) are the same. If we further define a state as the equally mixture of the above two cases, ${\rho _{mix}} = {{\left( {\rho _{ACB}^{\left(  +  \right)} + \rho _{ACB}^{\left(  -  \right)}} \right)} \mathord{\left/
 {\vphantom {{\left( {\rho _{ACB}^{\left(  +  \right)} + \rho _{ACB}^{\left(  -  \right)}} \right)} 2}} \right.
 \kern-\nulldelimiterspace} 2}$, then its covariance matrix will be $\gamma _{ACB}^{std}\left( {{\kappa _{11}},0,0,{\kappa _{22}}} \right)$. And from the sub-additivity of the secret key rate, we know $K\left( {{\rho _{mix}}} \right) \le {{\left( {K\left( {\rho _{ACB}^{\left(  +  \right)}} \right) + K\left( {\rho _{ACB}^{\left(  -  \right)}} \right)} \right)} \mathord{\left/
 {\vphantom {{\left( {K\left( {\rho _{ACB}^{\left(  +  \right)}} \right) + K\left( {\rho _{ACB}^{\left(  -  \right)}} \right)} \right)} 2}} \right.
 \kern-\nulldelimiterspace} 2} = K\left( {\rho _{ACB}^{\left(  +  \right)}} \right)$. Therefore, the lowest secret key rate case must happen at the condition ${\kappa _{12}} = {\kappa _{21}} = 0$. This further simplifies the searching process to two unknown parameters ${\kappa _{11}}$ and ${\kappa _{22}}$.

When ${\kappa _{12}} = {\kappa _{21}} = 0$,  it is found that the secret key rate is unchanged when $\left( {{\kappa _{11}},{\kappa _{22}},{\phi _x},{\phi _p}} \right)$  becomes $\left( {{-\kappa _{22}},{-\kappa _{11}},{\phi _p},{\phi _x}} \right)$. This means, if ${\phi _x} = {\phi _p}$, the lowest secret key rate happens at the condition ${\kappa _{11}} + {\kappa _{22}} = 0$, which further simplifies the searching process to only one unknown parameter ${\kappa _{11}}$.

Suppose $\Gamma  = \gamma _{ACB}^{std} + i\Omega $, and $\Gamma _{{i_1}{i_2}...{i_k}}^{{j_1}{j_2}...{j_k}}$ represents the minor determinant of order $k$ of $\Gamma $. The possible range for ${\kappa _{11}}$ and ${\kappa _{22}}$, limited by the uncertainty principle, are ${\kappa _{11}} \in \left[ {{{\bar \kappa }_{11}} - {R_x},{{\bar \kappa }_{11}} + {R_x}} \right]$ and ${\kappa _{22}} \in \left[ {{{\bar \kappa }_{22}} - {R_p},{{\bar \kappa }_{22}} + {R_p}} \right]$, where ${{\bar \kappa }_{11}} =  - {\phi _x}{{\Gamma _{124}^{234}} \mathord{\left/
 {\vphantom {{\Gamma _{124}^{234}} {\Gamma _{234}^{234}}}} \right.
 \kern-\nulldelimiterspace} {\Gamma _{234}^{234}}},{{\bar \kappa }_{22}} = {-\phi _p}{{\Gamma _{134}^{123}} \mathord{\left/
 {\vphantom {{\Gamma _{134}^{123}} {\Gamma _{134}^{134}}}} \right.
 \kern-\nulldelimiterspace} {\Gamma _{134}^{134}}}, $ and
\begin{equation*}
\begin{array}{l}
 {R_x} = {\left( {{V_B}{{\Gamma _{1234}^{1234}} \mathord{\left/
 {\vphantom {{\Gamma _{1234}^{1234}} {\Gamma _{234}^{234}}}} \right.
 \kern-\nulldelimiterspace} {\Gamma _{234}^{234}}} - {\phi _x}^2{{\left[ {\Gamma _{124}^{124}\Gamma _{234}^{234} - {{\left( {\Gamma _{124}^{234}} \right)}^2}} \right]} \mathord{\left/
 {\vphantom {{\left[ {\Gamma _{124}^{124}\Gamma _{234}^{234} - {{\left( {\Gamma _{124}^{234}} \right)}^2}} \right]} {{{\left( {\Gamma _{234}^{234}} \right)}^2}}}} \right.
 \kern-\nulldelimiterspace} {{{\left( {\Gamma _{234}^{234}} \right)}^2}}}} \right)^{1/2}} \\
 {R_p} = {\left( {{V_B}{{\Gamma _{1234}^{1234}} \mathord{\left/
 {\vphantom {{\Gamma _{1234}^{1234}} {\Gamma _{134}^{134}}}} \right.
 \kern-\nulldelimiterspace} {\Gamma _{134}^{134}}} - {\phi _p}^2{{\left[ {\Gamma _{134}^{134}\Gamma _{123}^{123} - {{\left( {\Gamma _{134}^{123}} \right)}^2}} \right]} \mathord{\left/
 {\vphantom {{\left[ {\Gamma _{134}^{134}\Gamma _{123}^{123} - {{\left( {\Gamma _{134}^{123}} \right)}^2}} \right]} {{{\left( {\Gamma _{134}^{134}} \right)}^2}}}} \right.
 \kern-\nulldelimiterspace} {{{\left( {\Gamma _{134}^{134}} \right)}^2}}}} \right)^{1/2}} \\
 \end{array}
 \end{equation*}

We note that, even if without these symmetry-induced simplifications, two facts indicate that the general linear searching algorithms also work effectively for the searching process: 1) the possible set $S_{\kappa}$ is a connected set; 2) the sub-additivity of the secret key rate indicates that usually there is only one minimum point of the secret key rate .

\subsection{Parameters for the numerical simulation}

\begin{figure}[t]
\centerline{\includegraphics[width=0.48\textwidth]{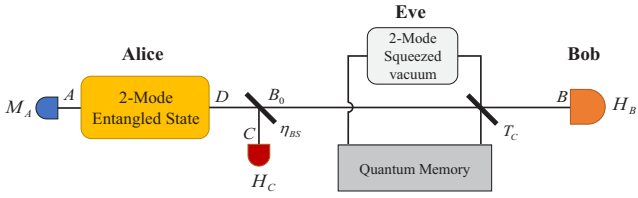}}
\caption{{\bf Schematic of the entangling cloner attack}. Eve generates a two-mode squeezed vacuum state, and send one mode to her quantum memory. She then interacts the other mode with mode ${B_0}$ using a beamsplitter (BS), whose transmittance equals to the channel transmittance ${T_C}$. She will send one mode after the BS to Bob, and keep the other mode in the quantum memory. After collecting enough many rounds, Eve will conduct the joint measurement on all states kept in her quantum memory. }\label{figS2}
\end{figure}

Here we explain the parameters used for the simulation in Fig. 3 of the main context. We consider the reverse reconciliation case, and assume that Bob uses homodyne detector. The reconciliation efficiency is assumed to be $0.95$. From the Fig.~\ref{figS1}, we know that generally speaking the smaller the ${V_A}$ is, the closer the ${\eta _A}$ approaches to $1$. However, if  ${V_A}$ is too small, its ability to tolerate the channel excess noise will decrease. Therefore, in the simulation, we choose $V_A = 3$ for $16$-QAM, and ${V_A} = 5$ for both 64-QAM and 256-QAM. The corresponding $\eta_A$ are $1-3.7\times {10^{-3}}, 1 - 1.4 \times {10^{ - 4}}$ and ${\eta _A} = 1 - 6.5 \times {10^{ - 8}}$, respectively.

To simulate the terms in the sub covariance matrix ${\gamma _{CB}}$, we assume the channel eavesdropping model is the entangling-cloner attack~\cite{Grosshans_QIC_2003}, which is commonly used in the performance simulation of one-way CV protocols. We note that the entangling-cloner attack may not be the optimal attack for QAM case. The reason we still choose it for the performance simulation is the channel usually behaves like this way in common experiments. The schematics of this attack is shown in Fig.~\ref{figS2}. Eve generates a two-mode squeezed vacuum state with variance $V_E = \left(1+T_C \epsilon_C \right) / \left(1-T_C\right)$, where $\epsilon_C$ is the channel excess noise and ${T_C}$ is the channel transmittance. She first sends one mode to her quantum memory, then interacts the other mode with mode ${B_0}$ using a beamsplitter (BS), whose transmittance equals to ${T_C}$. After this, she will send one mode after the BS to Bob, and keep the other mode in the quantum memory. After collecting enough many rounds, Eve will conduct a joint measurement on all states in her quantum memory.


For another parameter ${\eta _{BS}}$, which is the transmittance of the BS used to split the mode $D$ into modes $C$ and $B_0$. We choose it to be ${\eta _{BS}} = 0.9$. This means in our simulation, if $V_A = 3$, then the variance of ${\rho _{{B_0}}}$ is $2.8$, and if $V_A = 5$, then the variance of ${\rho _{{B_0}}}$ is $4.6$. Therefore, to compare $256$-QAM with ideal Gaussian case, we keep the variance of the state incident into the channel being the same, which means the variance for the ideal Gaussian case is also set to  $4.6$.
Additionally, we know that different $\eta_{BS}$ correspond to different entangled source, which will show different performances. And the higher the $\eta_{BS}$ is, the smaller the $V_A$ is. Thus, roughly speaking, higher $\eta_{BS}$ means better the performance. This parameter can also be optimized according to different channel conditions, if required.

\end{document}